Solubilization kinetics of oils by ionic and nonionic micelles: theoretical model

Alexey Kabalnov

Ink Splat Company, kabalnov@inkjet3D.com

**Abstract**

Experimental data on solubilization kinetics found in literature were analyzed by using the model proposed earlier (1). The rates of oil molecular exchange between the micellar core and the surrounding aqueous solution were determined. It was concluded that the solubilization of hydrocarbon molecules by nonionic micelles of ethylene oxide type is essentially barrier-free, that is, is diffusion controlled. It is quite different for ionic surfactants, where the rate is one-two orders of magnitude slower, indicating the existence of a potential barrier for hydrocarbons to get inside the micelles.
A Fickean diffusion model of solubilization has been proposed to explain these trends. For ionic micelles, the hydrocarbons are predicted to be excluded from the micellar double layer region because of their low dielectric constant. The Poisson-Boltzmann model was used to model this effect; the diffusion retardation factors were compared with the experiment and a fair agreement was seen. For nonionic groups, such as oligoethylene oxide, on the other hand, no such barrier was predicted to exist.
The analysis of this paper is performed only for the case of the 'slow' solubilization; it does not cover the case of the rapid, 'catastrophic' solubilization observed in the other group of experiments; the distinction between the slow and fast mechanisms is discussed and a possible explanation is suggested.

**Solubilization mechanisms**

As aqueous micelles have a hydrocarbon core, they are capable of solubilization of apolar materials such as hydrocarbons. From the pioneering work of Caroll (2), the kinetics of solubilization have been studied experimentally (3-7). The matter remains quite controversial and as is now being realized that the solubilization can proceed in two regimes: as a rapid process which we will call 'fast' or 'catastrophic' solubilization, and a much slower process, which we call 'slow' solubilization. The latter occurs closer to solubilization equilibrium, in the context made clearer below.
In the catastrophic solubilization process, a macroscopic oil drop can dissolve in a few minutes in presence of nonionic surfactants of the polyethylene oxide type (7). The radius of the oil drops decreases linearly with time and the rate of dissolution is approximately linear in the surfactant concentration. These 'fast' solubilization experiments are conducted in a single drop setup. The oil phase is not pre-saturated with the surfactant before the aqueous solution is added; in fact, it would be impossible to do as the surfactants of EO type are completely miscible with oils. Because of this, there must be a high incoming flux of the surfactant into the oil, which may cause a quick erosion of the particle, and this process should be considered as well; this aspect however has not yet been addressed theoretically.

The kinetics is completely different when the oil phase is added to the surfactant solution as a pre-made emulsion in water. In this case the solubilization is much slower, and it may take many hours for the oil drops to dissolve, even when they are much smaller in size (3, 4, 6). Naturally, in the pre-made emulsions, the oil is pre-saturated with the surfactant.

The mechanism of the catastrophic solubilization remains under investigation. It is argued that the process could be driven by the direct merge of the micelles coming from the aqueous phase with the oil drop (7). The difficulty with this argument is, it is not observed when the phases are pre-saturated with the surfactant, for example, in Ostwald ripening, where the direct merge of the micelles with the interface would have caused very high ripening rates which have not been observed experimentally (8, 9).

From this point on, we will focus this paper on the case of 'slow' solubilization, which occurs for the cases when the oil phase is pre-saturated with the surfactant, or when such pre-saturation is not necessary, as many ionic surfactants are virtually not soluble in oils.

**Theory of 'slow' solubilization kinetics: Dissolution of oil drops in presence of solubilizing micelles**

In aqueous micellar systems, (excluding the catastrophic solubilization scenarios) the oil cannot be absorbed by micelles directly, but only through the stage of molecular solution in water. In our previous paper (1), we analyzed coupling between the dissolution kinetics of a spherical oil drop in a micellar solution and the micellar dynamics, with the micelles constantly exchanging the oil with the medium. The equation below shows the predicted dissolution rate of an oil drop:

$$J = \frac{4\pi}{3}\frac{d}{dt}(r^3) = 4\pi r^2 D \Delta C \left(\kappa + \frac{1}{r}\right) \tag{1}$$

Here $D$ is the molecular diffusion coefficient of the oil in water, $\Delta C$ is the driving force of the particle dissolution, which is the difference in the oil concentration in water as a molecular solution at the surface of the oil drop and at the infinity. We use the concentration units reduced to the density of the oil, accordingly, $\Delta C$ is dimensionless, and $J$ has the dimension of cm³/s. The parameter $\kappa$ has the dimension of reciprocal length, which reflects on the oil molecular solution concentration profile around the dissolving particle. The value of $\kappa$ depends on the rate of molecular exchange between the oil in the aqueous solution and the micellar interior:

$$\kappa = \sqrt{\frac{\omega n_{mic}}{D}} \tag{2}$$

here $\omega$ is the source/sink term for the oil generation by a micelle [cm³/s], $n_{mic}$ is the number concentration of micelles per unit volume, $cm^{-3}$, and $D$ is the oil-in water molecular diffusion coefficient. More details of the physics of $w$ term will be given in the next section.

The value of $\kappa$ is often such that $\kappa >> 1/r$ and the eqn(1) can be simplified to:

$$\frac{4\pi}{3}\frac{d}{dt}(r^3) = -4\pi r^2 D\Delta C\kappa \tag{3}$$

or

$$\frac{dr}{dt} = -D\Delta C\kappa \tag{4}$$

In the solubilization kinetics experiment, $\Delta C = C_0 - C(t)$ is the difference between the aqueous molecular solubility of the oil $C_0$ and the transient aqueous concentration of the oil in water $C(t)$ which is in equilibrium with the micelles at the given moment of the solubilization process. As the micelles get saturated with the oil, the value of $C(t)$ increases and eventually reaches the value of $C_0$, when the mass transfer stops. Accordingly, the solubilization rate is expected to decrease with time as well, but at the initial stage of solubilization, $C(t)= 0$ and $\Delta C = C_0$ and :

$$\frac{dr}{dt}(t=0) = DC_0\kappa \tag{5}$$

Thus, the theory predicts that the dissolution kinetics of the oil drop is characterized by the linear decrease of the radius of the droplet, according to eqn (5), and the rate of decrease is proportional to the square root of the micellar concentration, according to eqn (2).

Other models of solubilization kinetics have been suggested. McClements and Dungan (3) has suggested an empirical equation for the solubilization kinetics, which covers not only initial rate of solubilization, but the whole kinetic range; however, the coefficients of this model are empirical parameters. Todorov et al (5) modified our model (1) by incorporating an energy barrier into the sink-source term for the micelle; the approach of this paper is similar to Todorov's with the difference that we are attempting to evaluate the value of this barrier, while in Todorov 's paper it is an empirical parameter.

We now proceed with the discussion of the micelle source sink term of Eqn 2 in more detail.

**Micelle source-sink term**

Assuming that there is no energy barrier for the oil molecule, the initial flux of the oil into an 'empty' 'oil-free' spherical micelle is equal to:

$$j = 4\pi r_{mic} DC_0 = 4\pi r_{mic} DC_0 \equiv \omega_0 C_0 \tag{6}$$

The equation is essentially the diffusional flux of the oil out the sphere of the micellar size; $r_{mic}$ is the micellar radius. The micelle is assumed to be spherical and the micellar size and shape changes with increasing the surfactant concentration are neglected.

The equation assumes that there is no barrier for the oil to get from the aqueous solution into the micellar interior and the process is diffusion controlled. However, this barrier can exist; to account for it one can add an Arrhenius factor to the equation:

$$\omega = \omega_0 exp\left(-\frac{\Delta G}{RT}\right) \equiv \frac{\omega_0}{F^*} \quad (7)$$

where $\Delta G$ is the activation free energy for the oil to enter or leave the micelle, and

$F^* = exp\left(\frac{\Delta G}{RT}\right)$ is the activation energy factor, whereas $\omega_0$ is the pre-exponent. We note that as oil diffuses into the micelle, the value of the energy barrier $\Delta G$ can change depending on the distance from the core of the micelle and some 'averaging' of the activation energy over the distance may be required. In this section, we will use the factor $F^*$, rather than the single exponential form; the detailed calculations of $F^*$ will be given in the following sections of this paper and Appendix.

Equation (7) the source/sink term for a single micelle; if the number of micelles per unit volume is $n_{mic}$, the total source-sink strength becomes:

$$\omega n_{mic} = \frac{4\pi r_{mic} n_{mic} D}{F^*} \quad (8)$$

or

$$\omega n_{mic} = \frac{3\phi_{mic} D}{F^* r_{mic}^2} \quad (9)$$

where $\phi_{mic}$ is the micellar volume fraction, or

$$\kappa = \frac{1}{r_{mic}} \sqrt{\frac{3\phi_{mic}}{F^*}} \quad (10)$$

**'Slow solubilization' experimental data: SDS and Tween 20**

McClements and Dungan, (3) and Ariyaprakai and Dungan (6), and Todorov et al (5) studied the solubilization of hydrocarbons by sodium dodecyl sulfate (SDS) and by Tween 20, a surfactant of ethylene oxide type. They observed nearly linear decreases in the particle radius with time. The dissolution rates did not depend on the initial radius of the particles [1]. Figure 1 shows the dissolution rates of the study (6) plotted vs square root of micellar volume fraction both for Tween 20 and for SDS; a good linearity is observed, and the slope of these lines can be used to evaluate the source-sink term for the micelles. The data are summarized in Table 1. The table provides the comparison of the sink-source terms $w$ with the theory; specifically, the ratio to the diffusion controlled based values $w_0$ is evaluated. Note that there are no adjustable parameters in the model. In that sense it is surprising that the model works so well for Tween 20 with the assumption of zero value of the energy barrier. However, it is not the case for SDS, where the diffusion rate is reduced by about one to two orders of magnitude over the pure diffusion control.

In the following section, we will be addressing specifically this issue, that is, the difference between the ionic and nonionic surfactants in terms of the easiness of the oil exchange with the micellar interior; this is the main topic of this paper.

---

[1] These data were in a striking contrast with the data of Pena and Miller (7), who found many orders of magnitude higher rates. Note that the first group experiments (3, 4, 6) was conducted in emulsion setup (that is, the oil phase was pre-equilibrated with the surfactant). The experiments of Todorov (5) were single drop experiments, but as SDS is virtually insoluble in hydrocarbons, it puts this experiment into the same category of the 'slow' solubilization.

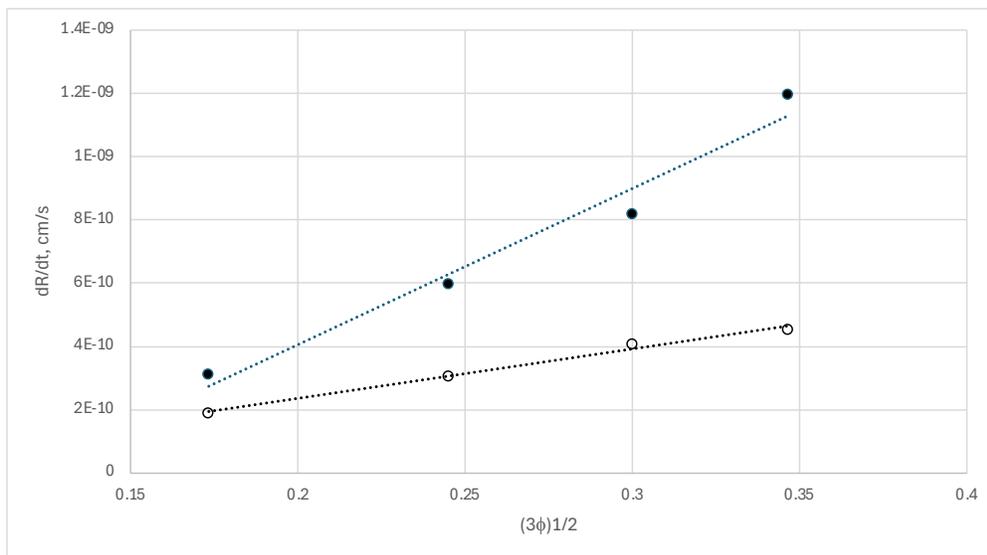
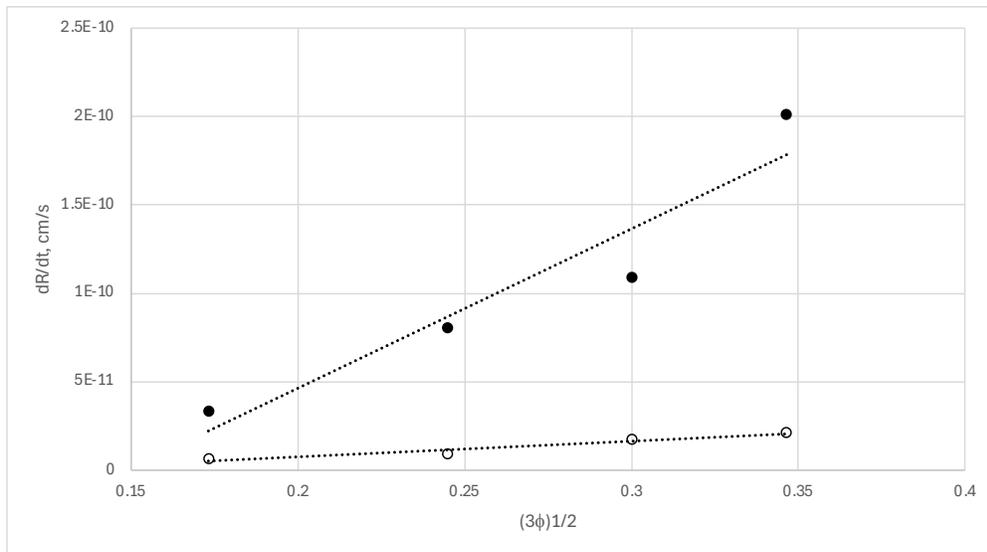

Figure 1. Initial solubilization rates of tetradecane (top) and hexadecane (bottom) in Tween 20 filled circles and SDS (open circles), plotted versus the square root of micellar volume fraction. The data are recalculated from Ref. (6), Figures 3 and 4.

Table 1. Parameters used in the model*

| Hydrocarbon | $C_{mol}$, ml/ml | D, cm²/s |
|---|---|---|
| n-decane | $7.1 \times 10^{-8}$ | $5.9 \times 10^{-6}$ |
| n-tetradecane | $3.7 \times 10^{-10}$ | $6.4 \times 10^{-6}$ |
| n-hexadecane | $2.7 \times 10^{-11}$ | $5.9 \times 10^{-6}$ |

Solubilities in water: Ref(10); Diffusivities of oils in water: Wilke-Cheng equation.

Table 2. Micelle source-sink rates presented as $F^*=w_0/w$ ratios over pure external diffusion control*.

| Surfactant | Hydrocarbon | Reference | Experimental setup | $F^*=w_o/w$ |
|---|---|---|---|---|
| SDS | decane | Ref (5), Fig 3 | single drop solubilization | 27 |
| SDS | decane | Ref (5), Fig 7 | single drop solubilization | 11.6 |
| SDS | tetradecane | Ref (6), Figure 4a | emulsion solubilization | 48 |
| SDS | hexadecane | Ref (6), Figure 3a | emulsion solubilization | 66 |
| SDS | tetradecane | Ref (6), Figure 7b | emulsion solubilization | 169 |
| SDS | hexadecane | Ref (6), Figure 7a | emulsion solubilization | 225 |
| Tween 20 | tetradecane | Ref (6), Figure 4b | emulsion solubilization | 2.2 |
| Tween 20 | hexadecane | Ref (6), Figure 3b | emulsion solubilization | 0.36 |
| Tween 20 | hexadecane | Ref 3 | emulsion solubilization | 2.6 |
| Tween 20 | hexadecane-to-octadecane | Ref 4 | composition ripening | 2.6 |

*Assumed micellar radii: 2.2 nm (SDS) and 3.2 nm (Tween 20). Note that in earlier work, McClements and Dungan (3, 4) suggested a higher value for the micellar radius of Tween 20 micelles, 5 nm, which was adjusted to 3.2 nm in the following publication. We are using this new value in all the calculations, to keep the consistency.

**Diffusion of oil inside the micelle: multilayer membrane 'onion' model**

We are now going to analyze the process by which oil is getting into a micelle. There are multiple steps to it. First, the oil needs to diffuse through the bulk of water to the micelle. For ionic micelles, it then needs to go through the diffuse layer of counter ions, which, in case of SDS, is the cloud of $Na^+$ ions. Then the layer of fixed charges of the Stern layer needs to be passed; for SDS it is the $SO_3^-$ non-dissociated $Na^+$ ion 'fence' attached to the carbon chain. Once these obstacles are passed, the oil gets into the hydrocarbon tail area. The diffusion can be modelled as the penetration through the three-layer stacked membrane, with the total diffusional resistance $R$ being the sum of the three terms:

$$\mathcal{R} = \mathcal{R}_w + \mathcal{R}_{dl} + \mathcal{R}_S \qquad (11)$$

The diffusional resistances are added together. Using the classical diffusion-through-membrane approach, each resistance is a function of the solute diffusion coefficient in it, the layer thickness and the partition coefficient:

$$\mathcal{R} = \frac{d_w}{D_w \Gamma_w} + \frac{d_{dl}}{D_{dl} \Gamma_{dl}} + \frac{d_S}{D_S \Gamma_S} \qquad (12)$$

we use the subscripts dl for the double layer, S for the Stern layer and w for the bulk water diffusion; $\Gamma$ are the partition coefficients for the hydrocarbon between the three layers. We choose the aqueous solution as the reference point, so the values of $\Gamma$ will be the ratios of the equilibrium concentrations in each layer to that in the bulk water, far away from the surface of the micelle; accordingly, $\Gamma_{water}$ is set at 1. The diffusion paths $d$ correspond to the different parts of the micelle, $d_w$ is the radius of the micelle, $d_{dl}$ is of the order of the Debye length of the double layer (more arguments about it below), and $d_S$ is the thickness of the Stern layer at the surface of the micelle. At this stage, we consider the micelles as a stack of flat layers; the conversion to spherical symmetry will be made shortly.

For nonionic micelles, the approach is similar, but only two layers of resistance can be identified; one being the bulk water, and the other being the 'brush' of polar heads, which in case of surfactants of ethylene oxide type is the polyethylene oxide brush. The brush molecules may contain from one to about twenty of ethylene oxide groups, depending on the surfactant; in aqueous solutions the polar groups are also heavily hydrated, at about 3-4 water molecules per one $CH_2CH_2O$ group.

$$\mathcal{R} = \mathcal{R}_w + \mathcal{R}_b = \frac{d_w}{D_w \Gamma_w} + \frac{d_b}{D_b \Gamma_b} \qquad (13)$$

The equations above are set for the planar geometry; we are more interested in the case of a spherical micelle. The same conceptual flow follows (Figure 2). The onion shells around the micelle add up to the overall diffusional resistance as:

$$\mathcal{R} = \sum \frac{\Delta r}{4\pi r^2 D \Gamma(r)} \qquad (14)$$

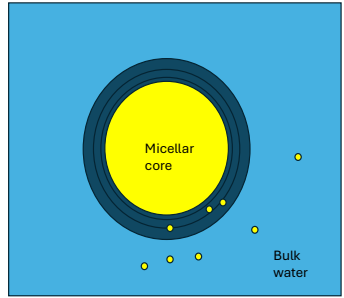

Figure 2. Cartoon explaining multilayer 'onion' diffusion model

This time we account for the fact that the membranes further away from the center contribute less to the resistance as they have a larger surface area. Replacing the sum with the integral, and assuming that the diffusion coefficient is not dependent on the distance from the micellar core, we get:

$$\mathcal{R} = \frac{1}{4\pi D} \int_{r_{mic}}^{\infty} \frac{dr}{r^2 \Gamma(r)} = \frac{1}{4\pi D} \int_{r_{mic}}^{\infty} \frac{1}{r^2} exp\left(\frac{\Delta G(r)}{k_B T}\right) dr \qquad (15)$$

We utilize the fact that the partition coefficient is due to the free energy difference for the solute in different locations around the micelle:

$$\Gamma(r) = exp\left(-\frac{\Delta G(r)}{k_B T}\right) \qquad (16)$$

Here the term $\Delta G(r)$ is the free energy of transfer of an oil molecule from the aqueous solution at infinity to a particular region at the micelle. The positive value of $\Delta G(r)$ indicates the existence of barrier and implies slowing down of the diffusion; this way, we account for the diffusional resistances coming form the different areas near the micelle (e.g. double layer, polyethylene oxide brush). Note that if there is no energy barrier, $\Delta G(r)=0$, the integral is equal to $\mathcal{R}_w = \frac{1}{4\pi D r_{mic}}$, which is the diffusional resistance of aqueous phase in the absence of any additional retardation factors.

Another aspect that needs to be discussed is the integration limits in Eqn 15. Ideally, the integration needs to be performed up to the region of the micellar hydrocarbon core and to include all the intermediate layers of the micelle, also known as 'palisade layers'. That is, the regions of polyethylene oxide brush and Stern layer must be included in the integral. If *ΔG(r)>0* in those locations, they add an extra diffusional resistance to the total. However, if *ΔG(r) <0*, they do not increase the rate of diffusion, but rather provide a small, potentially negligible addition to the diffusional resistance of the aqueous phase.

**Nonionic micelles of EO type**

For spherical micelles of oligoethylene oxide type, we could assume for simplicity that the ratio of heads-to tails is 1:1. Our focus is the comparison of the partition coefficients of the hydrocarbons in water and in the EO brush. Aliphatic hydrocarbons are known to be poorly soluble in both water and polyethylene oxide. However, they are considerably less insoluble in polyethylene oxide than in water. The data for liquid hydrocarbons seem to be missing, but they are available for hydrocarbon gasses, Ref (11). Figure 3 shows the effective 'partition coefficients' of methane, ethane and propane gases between PEG 400 and water; the partitioning 50 wt% PEG 400 is also included in the graph, to more closely model the hydrated surfactant brush. n-Propane is about 50x more soluble in PEG 400 than in water; in 50-50 mixture of PEG and water, it is still more soluble by about a factor of 2. Extending the trend in logarithmic coordinates, that is, assuming a constant free energy transfer per $CH_2$ group, one predicts that decane is more soluble in 50% PEG400 than in pure water by about 40X; the number for hexadecane is as high as 400. These partition coefficients are even larger for pure PEG400.

From this simple estimate, we conclude that for surfactants of polyethylene oxide type and linear alkanes, and fats, one can safely assume

$$\mathcal{R}_w \gg \mathcal{R}_b$$

$$\mathcal{R} \approx \mathcal{R}_w$$

and the incorporation of oil inside the micelle should proceed under external diffusion control, that is, every collision of an oil molecule that comes from the bulk with the micelle should result in the solubilization, as the brush provides no resistance to entrance; if anything, it attracts hydrocarbons to get in. Figure 4 shows similar data for the polypropylene oxide taken from Ref 12, for which these trends are only magnified. Nonionic surfactants of polysaccharide type is another class worth considering; no data is as readily available; the recently conducted measurements of the solubility of methane in water in presence of various saccharides, Ref (13) indicates a small increase in the solubility of methane in presence of sugars, which may mean that a similar effect may be observed for the surfactants with saccharide polar groups, for longer chain alkanes as well.

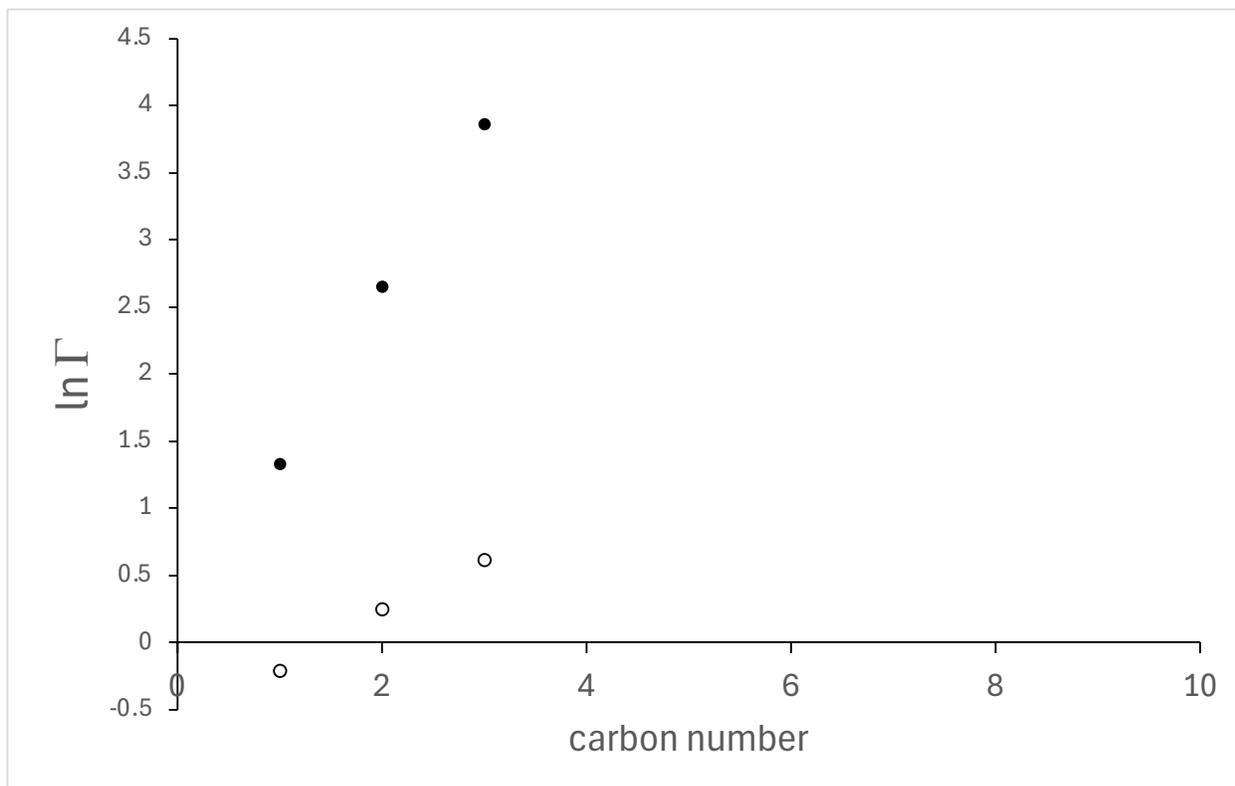

Figure 3. PEG400- water (filled circles) and ((50% PEG400 in water)-water, empty circles) effective 'partition coefficients' for methane, ethane and propane (Ref 11). The data recalculated from the values measured at 1 atmosphere pressure.

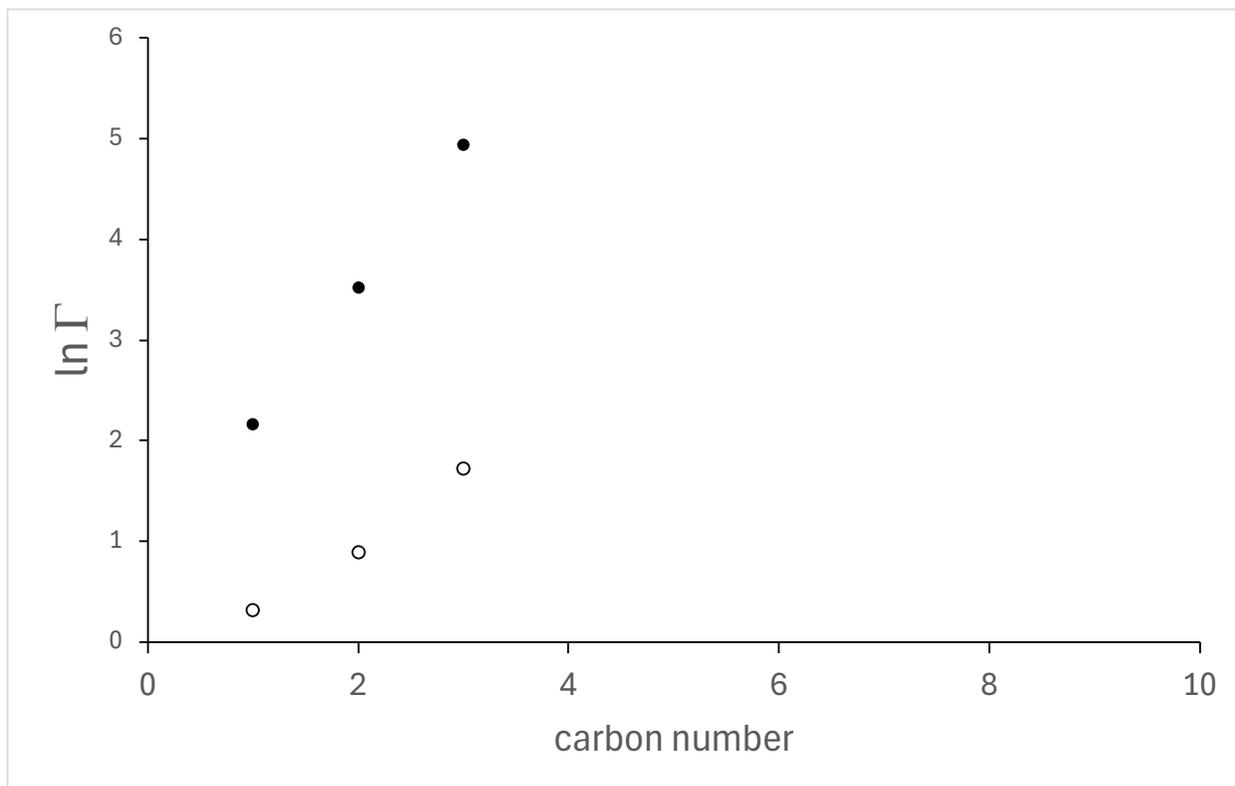

Figure 4. PPG400- water, filled circles, and (40% PPG400 in water)-water, empty circles, effective 'partition coefficients' for methane, ethane and propane (Ref 12). The data recalculated from the solubility values measured at 1 atmosphere pressure.

## Ionic micelles

We now transition to the case of ionic micelles. By contrast with EO chain, inorganic charged groups are expected to decrease the solubility of hydrocarbons in water, causing the local 'salting-out'. Thus, for n-hexane, adding 0.5 M of sodium sulfate to water decreases its solubility in water by about 7X , Ref (14). This alone implies that the charged groups must pose some barrier for hydrocarbons to enter the micelle. However, a stronger repulsive effect is expected in this case because of the charge separation in the double layer; we will proceed with the electrostatic energy barrier evaluation below.

The double layer can be viewed as a capacitor, filled with water as a dielectric medium (Figure 5); the energy of the capacitor is stored in an electric field, with the electrostatic energy density of:

$$u = \varepsilon_w \varepsilon_0 \frac{E_w^2}{2} \qquad (17)$$

We have the subscript $w$ for the field strength, indicating that the value is measured in water. When an oil molecule with molecular volume of $v_m$ and dielectric constant $\varepsilon_{oil} = 2$ comes in, the field strength locally increases as $\frac{\varepsilon_w}{\varepsilon_{oil}}$. We conclude that the energy penalty for an oil molecule to enter at a location $r$ and to replace the water that was located there is equal to:

$$\Delta G(r) = \frac{\varepsilon_0 \varepsilon_w}{2} \left(\frac{\varepsilon_w}{\varepsilon_{oil}} - 1\right) v_m E_w(r)^2 \qquad (18)$$

A numerical evaluation of this integral can be performed if the distribution of the electrical field strength around the micelle is known. The latter can be done by solving the Poisson Boltzmann equation of electrostatics for the micelle[2], Ref 15. Once this is done, the energy barrier function, Eqn 18, can be included in the diffusional resistance integral, eqn 15, and the diffusion retardation factor $F^*$ can be evaluated. [3]

Figure 6 shows the graph of the potential energy repulsion for decane molecule as it approaches an SDS micelle; for decane, the barrier of several $k_BT$ is predicted. Table 3 summarizes the values of the predicted diffusion retardation factors for different hydrocarbons. As expected, the predicted diffusion barrier increases with the molecular volume of hydrocarbons. The results of calculations are compared with the experimental values shown it Table 2. A fair agreement is observed, although the theory seems to underestimate the barrier somewhat. It is not clear at the moment if other factors such as the resistance of the Stern layer can account for this difference.

We also explored the effect of the assumed SDS micelle parameters on the diffusion retardation factor $F^*$ . Diminishing the Debye length around the micelle (e.g., by adding electrolytes) decreases the value of the factor slightly, see Table 4. On the other hand , the energy barrier is a strong function of the surface charge density. Accordingly, a strong dependence on the assumed degree of dissociation of the surface groups $\alpha$ is observed, see Table 5. The degree of dissociation of the SDS micelle surface groups is reported to be in 25 - 30% range (18); the best agreement with the experiment for diffusion retardation factor is seen if the value of 30% is assumed, which may be on the higher side of this range.

---

[2] Details of Poisson Boltzmann calculation for SDS micelle are given in Appendix 1.
[3] An explicit calculation can be performed for the planar geometry in case of a low charge density; see Appendix 2.

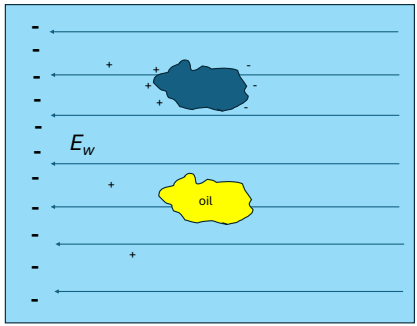

Figure 5. Cartoon explaining repulsion of low dielectric constant media from the double layer area

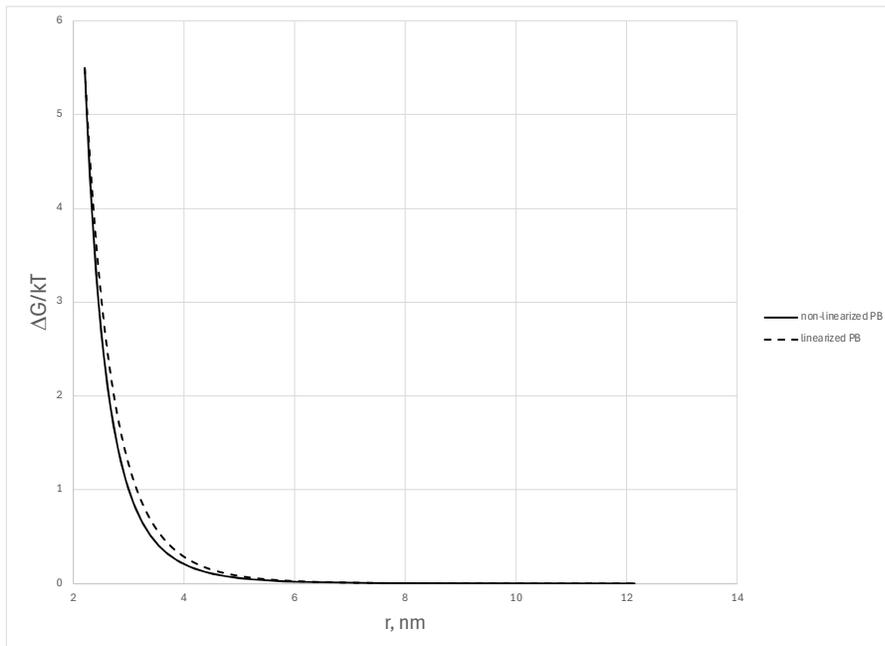

Figure 6. Electrostatic repulsion energy, $\Delta G(r)$ as function of distance from the center of the micelle for a spherical SDS micelle. The graph starts at the micellar surface at $r_{mic}$ = 2.2 nm. The other parameters of the model are : $1/\kappa_e$ = 3.11 nm; degree of micellar association N = 64, $\varepsilon_{oil}$ = 2, $\varepsilon_{water}$ = 81, the degree of dissociation of the polar heads $\alpha$ = 0.30. The diffusing hydrocarbon is decane, with the molar volume of 195 cm$^3$/mol. Both predictions of linearized and nonlinearized PB model are shown; the linearized model slightly overestimates the effect.

Table 3. Diffusion retardation factors $F^*=\mathcal{R}/\mathcal{R}_w$ for different hydrocarbons in SDS micelle. The parameters of the model are the same as in Figure 6.

|  | decane | dodecane | tetradecane | hexadecane | octadecane |
|---|---|---|---|---|---|
| $V_m$ | 195 | 229 | 256 | 294 | 328 |
| $F^* = \mathcal{R}/\mathcal{R}_w$, linearized PB model, $\alpha$= 0.3 | 12.19 | 25.75 | 48.02 | 119 | 274 |
| non-linearized PB | 10.19 | 21.21 | 39.24 | 97 | 221 |
| $F^*$, experiment | 12-27 |  | 48-159 | 66-225 |  |

Table 4. Diffusion retardation factors $\mathcal{R}/\mathcal{R}_w$ of decane for different values of Debye lengths, within non-linearized PB model, and $\alpha$ = 0.3. The rest of the parameters of the model are the same as in Figure 6.

| $1/\kappa_e$, nm | 3.11 | 2.2 | 1.1 |
|---|---|---|---|
| $\mathcal{R}/\mathcal{R}_w$ | 10.19 | 9.43 | 7.46 |

Table 5. Diffusion retardation factors $\mathcal{R}/\mathcal{R}_w$ of decane for different values of degree of polar group dissociation, within non-linearized PB model, and $1/\kappa_e$=3.11nm ( C=0.01 M).

| $\alpha$ | 0.25 | 0.27 | 0.30 |
|---|---|---|---|
| $\mathcal{R}/\mathcal{R}_w$ | 3.46 | 5.07 | 10.19 |

Conclusions and outlook

By analyzing literature data of solubilization kinetics of alkanes by Tween 20 and SDS micelles, the dynamics of micellar solubilization is extracted from the data. It is concluded that Tween 20 micelles absorb hydrocarbons without any energy barrier, whereas SDS micelles show some barrier, slowing down the kinetics by one-two orders of magnitude. The value of the barrier increases with the hydrocarbon chain length. A multilayer membrane 'onion' model is developed for the micelle solubilization, which explains the lack of barrier for surfactants of EO type; a similar behavior is expected for oligo (polypropylene oxide) and, possibly, even for saccharide polar heads. The prediction of the model is different for ionic micelles. The replacement of a high dielectric constant material (water) by a much lower dielectric constant material (oil) within the diffuse electrical double layer slows down the solubilization due to the electrostatic energy penalty.

A related topic to the solubilization kinetics is the transfer of the of the oil across the surfactant monolayer between emulsion drops, be it Ostwald ripening or composition ripening (16). Just like in the micelle solubilization case, no barrier by EO surfactants is predicted in emulsion case. For ionic surfactants some effect can be seen; however, as emulsion drops are bigger than micelles, the diffusional resistance of the bulk water phase is proportionally larger. Accordingly, the relative magnitude of the effect for emulsions will be smaller than for micelles.

## Appendix 1. Numerical solution of PB equation for SDS micelle.

The governing PB equation in spherical symmetry in presence of 1:1 electrolyte is (Ref. 15):

$$\frac{d^2\Phi}{dr^2} + \frac{2}{r}\frac{d\Phi}{dr} = \frac{en^*}{\varepsilon\varepsilon_0}\left[\exp\left(\frac{e\Phi}{k_BT}\right) - \exp\left(-\frac{e\Phi}{k_BT}\right)\right] \quad (19)$$

where $\Phi$ is the electrical potential, and $n^*$ is the electrolyte concentration in ions/m³ in infinity, far away from the micelle, $e$ is the electronic charge, $\varepsilon_0$ is the permittivity of the vacuum and $\varepsilon$ is the dielectric constant of water.

The first boundary condition for the equation is the field strength at the surface of the micelle, which is controlled by the electric charge density of the ionized groups $\sigma$:

$$E_w = \frac{d\Phi}{dr}(r = R_{mic}) = \frac{\sigma}{\varepsilon\varepsilon_0} \quad (20)$$

The second boundary condition is the fact that the electric potential at the infinity should tend to 0 far away from the micelle: $\Phi(r) \to 0$ at $r \to \infty$.

For small electrical potentials/weakly charged interfaces, the exponents can be expanded in series, yielding:

$$\frac{d^2\Phi}{dr^2} + \frac{2}{r}\frac{d\Phi}{dr} = \kappa_e^2 \Phi \quad (21)$$

where $\kappa_e = \sqrt{\frac{2n^*e^2}{\varepsilon\varepsilon_0 k_BT}}$ is the reciprocal Debye length. The linearized form of the equation can be solved analytically; the solution is shown below:

$$E_w(r) = \frac{Ze}{4\pi\varepsilon_w\varepsilon_0} \frac{(1+r\kappa_e)\exp[-\kappa_e(r-r_{mic})]}{r^2(1+\kappa_e r_{mic})} \quad (22)$$

The commonly used parameters for the SDS micelle are (17, 18): micellar radius $r_{mic}$ = 2.2 nm; the degree of micellar association $N$ = 64, the degree of dissociation of the polar heads $\alpha$ = 0.25- 0.3. As based on these values, the linearization approximation condition, $\frac{e\Phi}{k_BT} \ll 1$ should hold further from the micellar surface, but not at the very surface of the micelle, so the use of a non-linearized PB equation is warranted.

For the solution of PB equation, we utilized Runge-Kutta RK4 method (19) with the step of 0.002- 0.0005 nm; the integration was performed from the micellar radius, $r_{mic}$ = 2.2 nm to the total distance of $r$ = 12nm. In the process of iteration, the boundary condition of the field strength at the surface of the micelle $E_w(r_{mic})$ was held constant. The electrical potential $\Phi(r_{mic})$ was first set at the value of the linear approximation potential and then adjusted downwards till the condition of the potential tending to 0 at infinity was met, which was judged as the overlap of the numerical and the linearized potential curves at larger distances. Thus, for $1/\kappa_e$ = 2.2 nm (that is, the concentration of electrolyte of $C_{el}$= 0.02M) the values of the surface potentials were 64.56 mV and 59.50 mV for the linearized and non-linearized solutions, respectively. The typical graphs of the field strength and electrical potential vs distance are shown below in Figures 7 and 8.

It was determined that the potential barrier of diffusion is a strong function of the assumed degree of dissociation of ionic groups. On the other hand, the effect of the Debye length $1/\kappa_e$ is not as strong, see Tables 4 and 5  For SDS micelles,  the values covering the range from 0.25 to 0.30 are reported in literature, The best agreement for the retardation factors with the experiment was found at $\alpha$ = 0.30, which may be on the higher end of the values reported in literature, Ref (18).

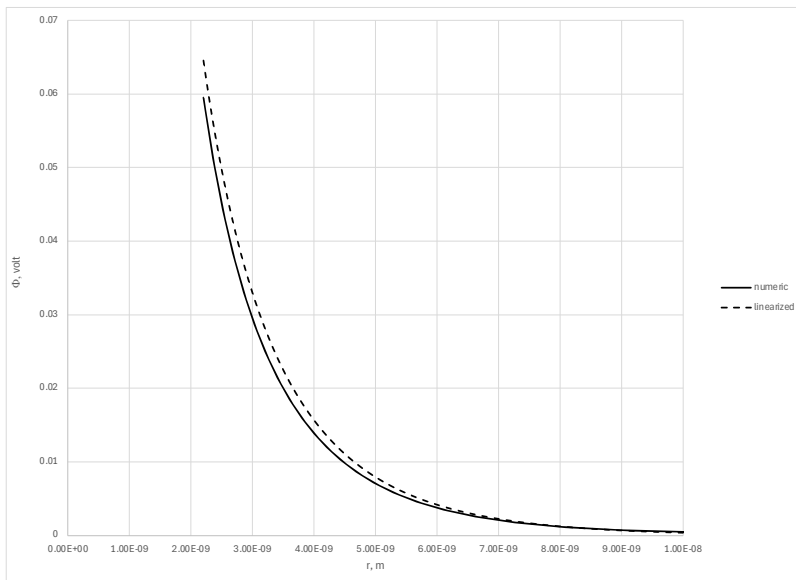

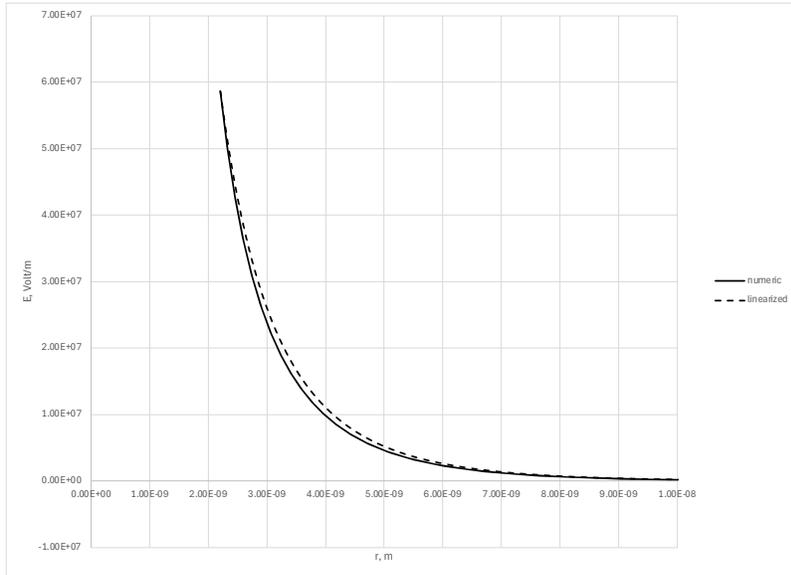

Figure 7. Dependence of the electrical potential and the field strength on the distance for SDS micelle. Micellar radius $r_{mic}$ = 2.2 nm; the degree of micellar association $N$ = 64, the degree of dissociation of the polar heads $\alpha$ = 0.25, $1/\kappa_e$ = 2.2 nm.

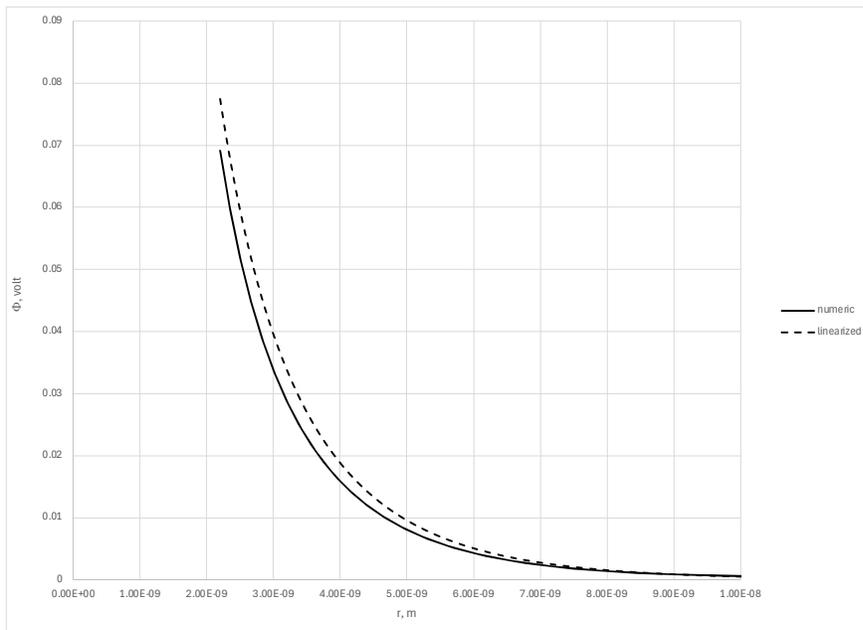

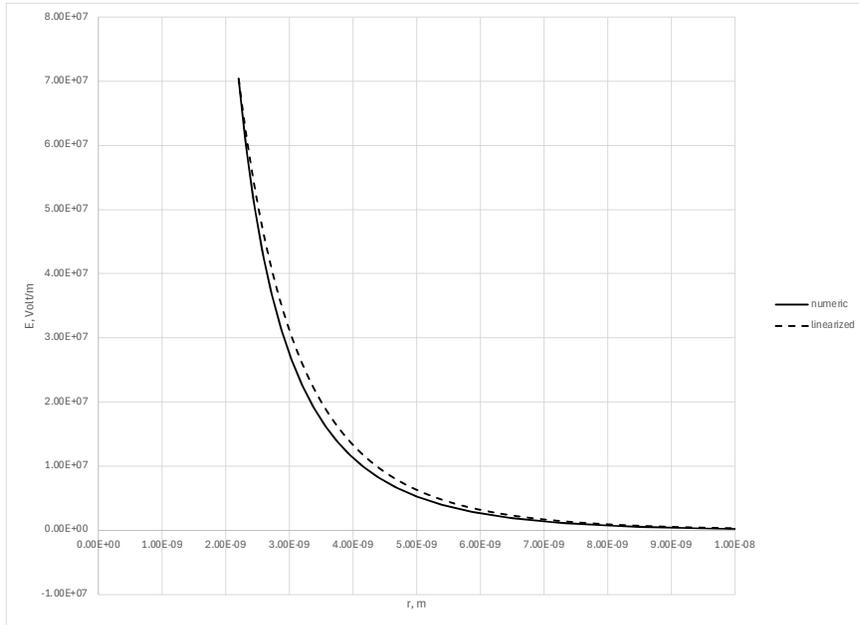

Figure 8. The same parameters as in Figure 7 were used except for $\alpha$ = 0.3. $\Phi_0$ = 0.0775 V (linearized) versus 0.06921 V (non-linearized) PB approximation.

## Appendix 2. Electrostatic penalty of bringing a dielectric into a flat weakly charged double layer

Consider an oil molecule in the vicinity of an electrical double layer (Figure 9). The double layer is considered to be flat, and only lightly charged. The aqueous vicinity of the double layer is modeled as a stack of flat membranes, with the diffusivity $D$, the thickness of each layer of $\Delta x$ and the partition coefficient between the location at distance $x$ from the interface, and pure water as $\Gamma$. For the stack of membranes, the diffusional resistance is equal to:

$$\mathcal{R} = \sum \frac{\Delta x}{D\Gamma(\mathrm{x})}$$

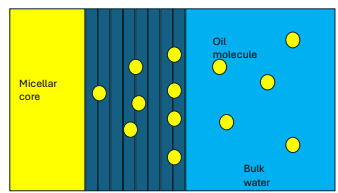

Figure 9. Cartoon illustrating diffusion of oil into the micelle in planar geometry.

The partition coefficient $\Gamma$ can be evaluated from the free energy of transfer between the layers and the pure water:

$$\mathcal{R} = \sum \frac{\Delta x}{D\Gamma} = \frac{1}{D}\int_0^L \exp\left(\frac{\Delta G(x)}{k_B T}\right) dx \qquad (23)$$

Assuming the energy barrier value is small compared to thermal energy $k_B T$, we expand the exponent in series:

$$\mathcal{R} = \frac{1}{D}\int_0^L \left(1 + \frac{\Delta G(x)}{k_B T}\right) dx = \frac{L}{D} + \frac{1}{Dk_B T}\int_0^L \Delta G(x) dx \qquad (24)$$

As we interested in the increment of the diffusion resistance over the pure water, which is $L/D$, the integral $\int_0^\infty \Delta G(x) dx$ is our primary interest; we put the integration limit to infinity as the value of the barrier will be decaying quickly (see below).

We assume the linearized Poisson-Boltzmann equation approximation, valid at a small surface charge densities, with the exponential decay with distance:

$$E_w = E_{0w} \exp(-\kappa_e x) \qquad (25)$$

where

$$E_{0w}(x=0) = \frac{\sigma}{\varepsilon_w \varepsilon_0} \qquad (26)$$

we arrive to:

$$\Delta G(x) = E_{0w}^2 \frac{\varepsilon_0 \varepsilon_w v_m}{2}\left(\frac{\varepsilon_w}{\varepsilon_{oil}} - 1\right) \exp(-2\kappa_e x) \qquad (27)$$

and

$$\Delta \mathcal{R} = \frac{1}{Dk_B T}\int_0^\infty \Delta G(x) dx = \frac{\sigma^2 v_m}{4\kappa_e \varepsilon_w \varepsilon_0 Dk_B T}\left(\frac{\varepsilon_w}{\varepsilon_{oil}} - 1\right) \qquad (28)$$

The diffusional resistance due to the double layer is then equal to:

$$\Delta \mathcal{R} = \frac{\lambda}{D} \frac{\sigma^2 v_m\left(\frac{\varepsilon_w}{\varepsilon_{oil}} - 1\right)}{4\varepsilon_w \varepsilon_0 Dk_B T} \qquad (29)$$

where $\lambda = 1/\kappa_e$ is the Debye length, and $\frac{1}{\Gamma} = \frac{\sigma^2 v_m\left(\frac{\varepsilon_w}{\varepsilon_{oil}} - 1\right)}{4\varepsilon_w \varepsilon_0 Dk_B T}$ is the reciprocal of the effective 'average partition coefficient' in that layer.